\input amstex
\input amsppt.sty            
\magnification=1200		
\pagewidth{6.  truein}
\pageheight{9. truein}
\hoffset = .15in
\voffset = 0in
\parskip=.1in
\NoBlackBoxes
\TagsOnRight

\def\q{\quad}
\def\p{\prime}
\def\d{\dot}
\def\dd{\ddot}

\topmatter
\headline={\hfill \the \pageno}
\title
The stability of cosmological scaling solutions
 \endtitle

\author
A. P. Billyard$^1$, A. A. Coley$^1$ and R. J. van den Hoogen$^2$ \endauthor
\affil  $^1$  Department of Mathematics,\\
Statistics and Computing Science\\
 Dalhousie University\\
        Halifax, Nova Scotia \enskip B3H 3J5\\
{}\\
$^2$ Department of Mathematics,\\ Statistics, and Computer Science\\
Saint Francis Xavier University\\
Antigonish, Nova Scotia  \enskip B2G 2W5
\endaffil
\endtopmatter

\subhead {Abstract} \endsubhead

We study the stability of cosmological scaling solutions within
the class of spatially homogeneous cosmological models with a perfect
fluid subject to the equation of state
$p_\gamma=(\gamma-1)\rho_\gamma$ (where $\gamma$ is a constant
satisfying $0<\gamma<2$) and a scalar field with an exponential
potential.  The scaling solutions, which are spatially flat isotropic
models in which the scalar field energy density tracks that of the
perfect fluid, are of physical interest.  For example, in these models
a significant fraction of the current energy density of the Universe
may be contained in the scalar field whose dynamical effects mimic
cold dark matter.  It is known that the scaling solutions are
late-time attractors (i.e., stable) in the subclass of flat isotropic models.
We find that the scaling solutions are stable (to shear and curvature
perturbations) in generic anisotropic Bianchi models when
$\gamma<2/3$.  However, when $\gamma>2/3$, and particularly for
realistic matter with $\gamma \ge 1$, the scaling solutions are
unstable; essentially they are unstable to curvature perturbations,
although they are stable to shear perturbations.  We briefly discuss
the physical consequences of these results.
  
\subhead {I. Introduction}\endsubhead 

Scalar field cosmological models are of great importance
in the study of the early Universe.  Models with a variety of
self-interaction potentials have been studied, and one potential that 
is commonly investigated and which arises in a number of physical
situations has an exponential dependence on the scalar
field [1 -- 5].  There have been a number of studies
of spatially homogeneous scalar field cosmological models
with an exponential potential, with particular emphasis
on the possible existence of inflation in such
models [1].

These models may also be important even if the 
exponential potential is too steep to drive inflation.  For example,
there exist `scaling solutions' in which the scalar field energy
density tracks that of the perfect fluid (so that at late
times neither field is neglible) [2].  In particular, in [3]
a phase-plane analysis of the spatially flat Friedmann-Robertson-Walker
(FRW) models showed that these scaling 
solutions are the unique late-time attractors whenever they 
exist.  The cosmological consequences of these scaling
models have been further studied in [4].  For example, in 
such models a significant fraction of the current energy
density of the Universe may be contained in the homogeneous scalar field
whose dynamical effects mimic cold dark
matter;  the tightest constraint on these cosmological models comes from 
primordial nucleosynthesis bounds on any such relic density [2 -- 4].

Clearly these scaling models are of potential
cosmological significance. It is consequently of prime importance
to determine the genericity of such models by studying their
stability in the context of more general spatially homogeneous 
models.  It is this question that we shall address in this paper.

\subhead{II. The Scaling Solution}\endsubhead 

The governing equations for a scalar field with an 
exponential potential
$$V = V_0 e^{-\kappa \phi}, \tag1$$
where $V_0$ and $\kappa$ are positive constants, evolving in a flat
FRW model containing a separately conserved perfect which satisfies
the barotropic equation of state 
$$p_\gamma = (\gamma -1) \rho_\gamma,$$
where the constant $\gamma$ satisfies $0\le\gamma\le 2$ (although we
shall only be interested in the range $0<\gamma < 2$ here), are given
by
$$\align 
\d{H} &= -\frac{1}{2} (\gamma \rho_\gamma + \d{\phi}^2), \tag2\\
\d{\rho}_\gamma &= -3 \gamma H \rho_\gamma, \tag3\\
\dd{\phi}&= -3H\d{\phi} + \kappa V, \tag4
\endalign 
$$
subject to the Friedmann constraint
$$H^2 = \frac{1}{3} (\rho_\gamma + \frac{1}{2} \dot{\phi}^2 + V),    \tag5 $$
where $H$ is the Hubble parameter, an overdot denotes ordinary
differentiation with respect to time $t$, and units have been chosen
so that $8 \pi G =1$.  We note that the
total energy density of the scalar field is given by
$$\rho_\phi = \frac{1}{2}\dot{\phi}^2 +V. \tag6$$ 

Defining
$$x \equiv \frac{\dot{\phi}}{\sqrt{6}H} \q , \q y \equiv \frac{\sqrt{V}}{\sqrt{3}H},  \tag7 $$
and the new logarithmic time variable $\tau$ by
$$\frac{d \tau}{dt} \equiv H, \tag8$$
equations (2) -- (4) can be written as the plane-autonomous system [3]:
$$
\align
x^\p & = -3x + \sqrt{\frac{3}{2}} \kappa y^2 + \frac{3}{2} x [2x^2 + \gamma (1-x^2 -y^2)], \tag9\\
y^\p &= \frac{3}{2}y \left[-\sqrt{\frac{2}{3}} \kappa x + 2x^2 + \gamma (1-x^2 -y^2)\right], \tag10
\endalign$$
where a prime denotes differentiation with respect to $\tau$, and 
equation (5) becomes
$$\Omega + \Omega_\phi =1, \tag11$$
where
$$\Omega \equiv \frac{\rho_\gamma}{3 H^2}, \q \Omega_\phi \equiv \frac{\rho_\phi}{3H^2} = x^2 + y^2, \tag12$$
which implies that $0 \leq x^2 +y^2 \leq 1$ for $\Omega \geq 0$ so that the
phase-space is bounded.

A qualitative analysis of this plane-autonomous system
is given in [3].  The well-known power-law inflationary
solution for $\kappa^2<2$ [1] corresponds to the equilibrium point
$x = \kappa /\sqrt{6}$, $y = (1 - \kappa^2/6)^{1/2}$ ($\Omega_\phi =1$, $\Omega =0$) of the system (9)/(10), which is shown to be
stable (i.e., attracting) for $\kappa^2 < 3 \gamma$ in the presence of a
barotropic fluid.  Previous analysis has shown that when $\kappa^2<2$
this power-law inflationary solution is a global attractor in 
spatially homogeneous models in the absence of a perfect 
fluid (except for a subclass of Bianchi type IX models 
which recollapse).

In addition, for $\gamma > 0$ there exists a scaling solution
corresponding to the equilibrium point
$$x = x_0 = \sqrt{\frac{3}{2}} \frac{\gamma}{\kappa}, \q y = y_0 = [3
(2 - \gamma) \gamma/2 \kappa^2]^{\frac{1}{2}}, \tag13$$ whenever
$\kappa^2 > 3\gamma$.  The linearization of system (9)/(10)
about the equilibrium point (13) yields the two 
eigenvalues with negative real parts
$$-\frac{3}{4}\left( 2-\gamma\right) \pm \frac{3}{4\kappa} \sqrt{(2-\gamma)
[24\gamma^2-\kappa^2(9\gamma-2)]} \tag14$$
when $\gamma<2$.  The equilibrium point is consequently stable (a
spiral for $\kappa^2 > 24 \gamma^2/(9\gamma-2)$, else a node) so that
the corresponding cosmological solution is a late-time attractor in
the class of flat FRW models in which neither the scalar-field nor the
perfect fluid dominates the evolution.  The effective equation of
state for the scalar field is given by
$$\gamma_\phi \equiv \frac{(\rho_\phi + p_\phi)}{ \rho_\phi} = \frac{2x^2_0}{x^2_0 + y^2_0} = \gamma, $$
which is the same as the equation of state parameter for the
perfect fluid.  The solution is referred to as a scaling solution since the energy
density of the scalar field remains proportional to that of 
the barotropic perfect fluid according to
$\Omega/\Omega_\phi = \kappa^2/3\gamma -1$ [2].  Since the
scaling solution corresponds to an equilibrium point of the system
(9)/(10) we note that it is a self-similar cosmological model [6].

\subhead{III. Stability of the Scaling Solution}\endsubhead

Let us study the stability of the scaling solution 
with respect to anisotropic and curvature perturbations within
the class of spatially homogeneous models.

\subhead{A.  Bianchi I models}  \endsubhead

In order to study the stability of the 
scaling solution with respect to shear perturbations we shall first 
investigate the class of anisotropic Bianchi I models, which 
are the simplest spatially homogeneous generalizations
of the flat FRW models which have non-zero 
shear but zero three-curvature.  The governing equations in the Bianchi I 
models are 
equations (3) and (4), and equation (5) becomes
$$H^2 = \frac{1}{3} \left(\rho_\gamma + \frac{1}{2} \d{\phi}^2 + V\right) + \Sigma^2, \tag15$$
where $\Sigma^2 = \frac{1}{3} \Sigma^2_0 R^{-6}$ is the contribution due to the shear, where $\Sigma_0$ is a constant and $R$ is the  
scale factor.  Equation (2) is replaced by the time derivative of equation (15).

Using the definitions (7), (8) and (12) we can
deduce the governing ordinary differential equations. Due to 
the $\Sigma^2$ term in (15) we can no longer use this equation
to substitute for $\rho_\gamma$ in the remaining equations, and 
we consequently obtain the three-dimensional autonomous 
system:
$$\align
x^\p &= -3x + \sqrt{\frac{3}{2}} \kappa y^2 + \frac{3}{2} x [2 + (\gamma -2) \Omega -2y^2], \tag16\\
y^\p &= \frac{3}{2}y \left\{ - \sqrt{\frac{2}{3}} \kappa x +2 + (\gamma -2) \Omega
-2y^2\right\}, \tag17\\
\Omega^\p &= 3 \Omega \{(\gamma -2)(\Omega -1) -2y^2\}, \tag18
\endalign $$
where equation (15) yields
$$1 - \Omega -x^2 -y^2 = \Sigma^2 H^{-2} \geq 0, \tag19$$
so that we again have a bounded phase-space.

The scaling solution, corresponding to the flat FRW 
solution, is now represented by the equilibrium point
$$x = x_0, ~~y = y_0, ~~\Omega =1 - \frac{3 \gamma}{\kappa^2}. \tag20$$
The linearization of system (16) -- (18) about the equilibrium 
point (20) yields three eigenvalues, two of which are given by (14)
and the third has the value $-3(2-\gamma)$, all with negative real
parts when $\gamma<2$.  Consequently the scaling solution is stable to 
Bianchi type I shear perturbations.

\subhead{B.  Curved FRW models}  \endsubhead

In order to study the stability of the scaling solution
with respect to curvature perturbations we shall first study the
class of FRW models which have curvature but no shear. Again
equations (3) and (4) are valid, but in this case equation (5) becomes
$$H^2 = \frac{1}{3} (\rho_\gamma + \frac{1}{2} \d{\phi}^2 +V) + K, \tag21$$
where $K = -k R^{-2}$ and $k$ is a constant that can be scaled to $0$, $\pm 1$.
Equation (2) is again replaced by the time derivative of equation
(21).

As in the previous case we cannot use equation (21) to replace $\rho_\gamma$, and using the definitions
(7), (8) and (12) we 
obtain the three-dimensional autonomous system:
$$\align
x^\p &= -3x + \sqrt{\frac{3}{2}}\kappa y^2 + \frac{3}{2}x \left[ \left( \gamma - \frac{2}{3} \right) \Omega + \frac{2}{3} (1 + 2x^2 -y^2)  \right], \tag22\\
y^\p &= \frac{3}{2} y \left\{ -\sqrt{\frac{2}{3}} \kappa x + \left( \gamma -\frac{2}{3}  \right) \Omega + \frac{2}{3} (1+2x^2 -y^2) \right\}, \tag23\\
\Omega^\p &= 3 \Omega \left\{\left(\gamma - \frac{2}{3}\right)(\Omega -1) + \frac{2}{3} (2x^2 -y^2)  \right\}, \tag24
\endalign$$
where
$$1 - \Omega - x^2 -y^2 = KH^{-2}. \tag25$$
The phase-space is bounded for $k =0$ or $k=-1$, but not for $k = +1.$

The scaling solution again corresponds to the equilibrium point (20).
The linearization of system (22) -- (24) about this equilibrium point
yields the two eigenvalues with negative real parts given by (14) and
the eigenvalue $(3 \gamma -2)$. Hence the scaling solution is only
stable for $\gamma < \frac{2}{3}$.  For $\gamma > \frac{2}{3}$ the
equilibrium point (20) is a saddle with a two-dimensional stable
manifold and a one-dimensional unstable manifold.

Consequently the scaling solution is unstable
to curvature perturbations in the case of realistic
matter $(\gamma \geq 1)$; i.e., the scaling solution is no longer
a late-time attractor in this case.  However, the scaling solution
does correspond to an equilibrium point of the governing 
autonomous system of ordinary differential equations
and hence there are cosmological models that can spend
an arbitrarily long time `close' to this solution.  Moreover,
since the curvature of the Universe is presently constrained
to be small by cosmological observations, it is possible
that the scaling solution could be important in 
the description of our actual Universe.  That is, not enough
time has yet elapsed for the curvature instability to
have effected an appreciable deviation from the flat
FRW model (as in the case of the standard perfect fluid
FRW model).

Hence the scaling solution may still be of physical interest.  To
further study its significance it is important to determine its
stability in a general class of spatially homogeneous models.  We
shall therefore study the stability of the scaling solution in the
(general) class of Bianchi type VII$_h$ models, which are perhaps the most
physically relevant models since they can be regarded as
generalizations of the open (negative-curvature) FRW models.

\subhead{C. Bianchi VII$_h$ models}  \endsubhead

The Bianchi VII$_h$ models are sufficiently complicated 
that a simple coordinate approach
(similar to that given above) is not desirable.
To study Bianchi VII$_h$ spatially homogeneous 
models with a minimally coupled scalar field 
with an exponential
potential and a barotropic perfect fluid it is best
to employ a group-invariant orthonormal frame approach
with expansion-normalized state variables governed by
a set of dimensionless evolution equations (constituting
a `reduced' dynamical system) with respect to a 
dimensionless time subject to a non-linear constraint
[6], generalizing previous work in which there is 
no scalar field [7] and in which there is no matter [8].

The reduced dynamical system is seven-dimensional (subject to a
constraint) [9].  The scaling solution is again an equilibrium point
of this seven-dimensional system. This equilibrium point, which only
exists for $\kappa^2>3\gamma$, has two eigenvalues given by (14) which
have negative real parts for $\gamma<2$, two eigenvalues
(corresponding to the shear modes) proportional to $(\gamma -2)$ which
are also negative for $\gamma <2$, and two eigenvalues (essentially
corresponding to curvature modes) proportional to $(3 \gamma -2)$
which are negative for $\gamma <\frac{2}{3}$ and positive for $\gamma
>\frac{2}{3}$ [9].  The remaining eigenvalue (which also corresponds
to a curvature mode) is equal to $3\gamma-4$.  Hence for $\gamma <
\frac{2}{3}$ ($\kappa^2 > 3 \gamma$) the scaling solution is again
stable.  However, for realistic matter ($\gamma \geq 1$) the
corresponding equilibrium point is a saddle with a (lower)
four- or five-dimensional stable manifold (depending upon whether 
$\gamma>4/3$ or $\gamma<4/3$, respectively).

\subhead{IV. Discussion} \endsubhead

Perhaps these stability results can be understood heuristically as 
follows.  From the conservation law the barotropic matter redshifts as 
$R^{-3\gamma}$.  In subsection III.A we saw that in this case the shear 
$\Sigma^2$ redshifts as $R^{-6}$ and so always redshifts faster than the 
matter, resulting in the stability of the scaling solution.  We note that 
the bifurcation that occurs at $\gamma=2/3$ in subsection III.B 
corresponds to the case in which the curvature $K$ is formally equivalent 
to a barotropic fluid with $\gamma=2/3$, and in which both the matter and 
the curvature redshift as $R^{-2}$.  For $\gamma>2/3$, the barotropic 
matter redshifts faster than $R^{-2}$ and the curvature eventually 
dominates.  A complete qualitative analysis of cosmological models with a 
perfect fluid and a scalar field with an exponential potential will be 
undertaken in future work [cf. 9].
 
\heading{REFERENCES}\endheading

\roster

\item"[1]." J.J. Halliwell, Phys. Lett. B{\bf 185}, 341 (1987); A.B. Burd
and J.D. Barrow, Nucl. Phys. B{\bf 308}, 929 (1988); Y. Kitada and K. Maeda, Class. Quantum 
Grav. {\bf 10}, 703, (1993); A.A. Coley, J. Iba\~nez and R.J. van den Hoogen, J. Math. Phys. {\bf 38}, 5256 (1997).

\item"[2]." C. Wetterich, Nucl. Phys. B{\bf 302}, 668 (1988); D. Wands, E.J. 
Copeland and A.R. Liddle, Ann. N.Y. Acad. Sci. {\bf 688}, 647 (1993).

\item"[3]." E.J. Copeland, A.R. Liddle and D. Wands, Phys. Rev. D{\bf 57}, 4686 (1998).

\item"[4]." P.G. Ferreira and M. Joyce, Phys. Rev. Lett. {\bf 79},
4740 (1997) and Phys. Rev. D{\bf58}, 023503 (1998); C. Wetterich,
Astron. Astrophys. {\bf 301}, 321 (1995).

\item"[5]." E.J. Copeland, A. Lahiri and D. Wands, Phys. Rev D{\bf 50}, 4868 (1994).

\item"[6]." J. Wainwright and G.F.R. Ellis, {\it Dynamical Systems in Cosmology}, Cambridge University Press (1997).

\item"[7]." C.G. Hewitt and J. Wainwright, Class. Quantum Grav.
{\bf 10}, 99 (1993).

\item"[8]." R.J. van den Hoogen, A.A. Coley and J. Iba\~nez, Phys. Rev. D{\bf 55}, 5215, (1997).

\item"[9]." A.P. Billyard, A.A. Coley and R.J. van den Hoogen, in preparation (1998).

\endroster
 
\enddocument